\documentclass[12pt,a4paper]{article}
\usepackage{graphicx}
\usepackage{authblk}
\usepackage{amsfonts}
\usepackage[dvips]{color}

\newcommand{\beq}{\begin{equation}}
\newcommand{\eeq}{\end{equation}}
\newcommand{\beqa}{\begin{eqnarray}}
\newcommand{\eeqa}{\end{eqnarray}}

\def\ket#1{|\,#1\,\rangle}
\def\bra#1{\langle\, #1\,|}

\def\proj#1#2{\ket{#1}\bra{#2}}

\newcommand{\pgen}{\mathbb{P}}
\newcommand{\upol}{\hat{U}_{{\rm pol}}}
\newcommand{\rop}{\hat{\rho}}
\newcommand{\sop}{\hat{\sigma}}

\begin{document}

\title{Entanglement versus quantum degree of polarization}
\author[]{Iulia Ghiu\thanks{email: iulia.ghiu@g.unibuc.ro}}
\affil[]{University of Bucharest, Faculty of Physics, Centre for Advanced Quantum Physics, PO Box MG-11, R-077125, Bucharest-Magurele, Romania }

\maketitle

\begin{abstract}
In this article we make a comparison between the behavior of entanglement and quantum degree of polarization for a special class of states of the radiation field, namely the Bell-type diagonal mixed states. For the three-photon mixed states we have plotted the concurrence and the Chernoff quantum degree of polarization in terms of the parameter, which defines the state. We find the the entanglement and the quantum degree of polarization are incomparable measures.
\end{abstract}

\section{Introduction}

Entanglement is the main ingredient in most applications of quantum information theory \cite{Nielsenbook, Sen}. For example, it is widely used in quantum cryptography or for obtaining the mutually unbiased bases required for quantum key distribution \cite{Ghiu-2012, Ghiu-2013, Ghiu-math-2014, Ghiu-phys-2014, Manko-1-2014, Manko-2-2014}, in quantum teleportation, or in superdense coding \cite{Sen}. The resources based on non-Gaussian states have attracted a lot of attention due to the fact that they are more efficient for some quantum processes such as teleportation \cite{Olivares} or cloning \cite{Cerf}. Recently, a new measure of non-Gaussianity was introduced and investigated \cite{Marian-2013}, \cite{Marian-pra-2013}. During the last years, two new properties of quantum states were defined: quantum discord \cite{Modi} and  Einstein-Podolsky-Rosen steerability \cite{Scarani}, which are different from entanglement and Bell nonlocality. The behavior of discord and steerability under the influence of a thermal bath was discussed in Refs. \cite{Isar-2015, Isar-eur-2017, Isar-proc-2017, Serban-2015, Isar-2017}.

Another important feature of the radiation field is the polarization. While in classical optics, there is the well-known definition based on the Stokes parameters, in quantum optics there are many proposals for introducing the degree of polarization. A choice for evaluating the quantum degree of polarization is the use of distance-type measures, such as Hilbert-Schmidt, Bures \cite{Bjork-2005}, or Chernoff bound \cite{Ghiu-2010, Bjork-2010}.

An investigation of the entanglement and polarization was done for photon-added coherent states in Ref. \cite{Silva}. The task of this paper is to make a comparison between the entanglement, using the concurrence, and polarization based on the quantum Chernoff bound of two-mode three-photon states, which are diagonal in the Bell-type basis. A different analysis was done by us by using the Hilbert-Schmidt and Bures measures for the quantum degree of polarization \cite{Ghiu-capitol}.

The paper is organized as follows. Section 2 presents a brief review of the quantum Chernoff bound, as well as the quantum degree of polarization based on this quantity. The main results of this article are given in Sec. 3, where the three-photon Bell-type diagonal mixed states are analyzed in detail with the help of the Chernoff degree of polarization and concurrence. Further, two particular cases are investigated: the Werner-type state and the Bell-type diagonal state depending on a parameter, for which we have plotted both the concurrence and the Chernoff degree of polarization. Our conclusions are outlined in Sec. 4, where we emphasize that
the entanglement and the quantum degree of polarization are incomparable measures.

\section{Chernoff quantum degree of polarization: a brief review}

The problem of discriminating of two probability distributions was investigated in the asymptotic limit by Chernoff in 1952 \cite{Chernoff}, who found an upper bound on the minimal error probability. This bound is known as the classical Chernoff bound and has many applications in statistical decision theory. Its generalization to the quantum case was recently proved by Nussbaum and Szko\l a  \cite{Nussbaum} and Audenaert {\it et al.} \cite{Acin}, and analyzes the following scenario: $k$ identical copies of a quantum system
are prepared in the same unknown state, which is either $\hat \rho$ or
$\hat \sigma$. Then, one has to determine the minimal probability of error
by testing the copies in order to draw a conclusion about the identity
of the state. The minimal error probability of discriminating two equiprobable states in a measurement performed
on $k$ independent copies is \cite{Kargin, Auden}
\begin{equation}
P_{\rm min}^{(k)}(\rop,\, \hat \zeta)=\frac{1}{2}\left(1-\frac{1}{2}
||{\rop }^{\otimes k}-{\hat \zeta}^{\otimes k}||_1\right)
\label{error},
\end{equation}
where $||\hat A||_1:=\mbox{Tr}\sqrt{\hat A^{\dagger} \hat A}$ is
the trace norm of a trace-class operator $\hat A$.

Due to the fact that the minimal error probability given by Eq. (\ref{error}) is in general difficult to be computed, one has to consider the asymptotic limit. For a large number of identical copies, an upper bound of the minimal probability of error (\ref{error}) has the expression \cite{Acin}:
$$P^{(k)}_{\rm min}(\rop ,\, \hat \zeta)\sim \exp \left[-k\; \xi _{QCB}(\rop , \,\hat \zeta)\right]. $$

The positive quantity  $\xi_{QCB}(\rop,\, \hat \zeta)$ is called the quantum Chernoff bound \cite{Nussbaum, Acin} and is defined as follows
\begin{equation}
\xi_{QCB}(\rop ,\, \hat \zeta):= \lim_{k\to \infty }-\frac{\ln P^{(k)}_{\rm min}(\rop ,\, \hat \zeta)}{k}
=-\ln \left[ \min_{s\in [0,1]}\mbox{Tr}\left({\rop }^s{\hat \zeta}^{1-s}\right) \right].
\label{xi}
\end{equation}

 Let us consider a mixed two-mode state $\rop $ of the quantum radiation field. Further, we use the notation $\ket{k,N-k}=\ket{k}_H\otimes \ket{N-k}_V$, with $k=0$, 1,..., $N$ for the standard two-mode Fock basis, where $H$ means horizontal polarization, while $V$ the vertical one.
 The class of linear polarization transformations is a group of unitary operators $\upol $
on the two-mode Hilbert space ${\cal H}_{H}\otimes {\cal H}_{V}$, these transformations being constructed with the help of the Stokes operators:
\beqa
\hat S_1&:=&\hat a _{H}^\dagger \hat a_{V}+\hat a _{H} \hat a_{V}^\dagger , \nonumber \\
\hat S_2&:=&\frac{1}{i} \left( \hat a _{H}^\dagger \hat a_{V}
-\hat a _{H} \hat a_{V}^\dagger \right), \nonumber \\
\hat S_3&:=&\hat a _{H}^\dagger \hat a_{H}-\hat a _{V}^\dagger \hat a_{V}. \nonumber
\eeqa

The operators ${\hat U}_{\rm pol}$ can be parametrized in terms of the Euler angles  $\phi $, $\theta $, $\psi $ as follows:
\beq
{\hat U}_{\rm pol}(\phi,\,\theta,\,\psi)=\exp{\left(-i\, \frac{\phi}{2}\, \hat S_3\right)}
\exp{\left(-i\, \frac{\theta}{2}\, \hat S_2\right)} \exp{\left(-i\, \frac{\psi}{2}\, \hat S_3\right)}.
\label{Upol}
\eeq

A state $\sop $ that remains invariant under any polarization transformation (\ref{Upol}), {\it i.e.} $\upol \sop \upol^\dagger =\sop $, is unpolarized \cite{Ghiu-2010}, its spectral decomposition being \cite{Lehner}
\begin{equation}
\sop =\sum_{N=0}^\infty \; \pi_N \frac{1}{N+1}\, {\hat P}_N,
\label{sigma}
\end{equation}
where
$
{\hat P}_N:=\sum_{n=0}^N |n,N-n \rangle \langle n,N-n|
$
is the projection operator onto the vector subspace of the $N$-photon states.
The parameters $\pi_N$ satisfy the normalization condition
$
\sum_{N=0}^{\infty}\pi_N=1.
$

The Chernoff quantum degree of polarization of a two-mode state of the quantum radiation field is defined as follows \cite{Ghiu-2010,   Bjork-2010}:
\begin{equation}
\mathbb{P}_C(\hat \rho):=1-\max_{\sop \in {\cal U}}\left[ \min_{s\in [0,1]}
\mbox{Tr}({\hat \rho}^s{\hat \sigma}^{1-s})\right],
\label{mas-c}
\end{equation}
where the maximization is evaluated over the set ${\cal U}$ of unpolarized states $\sop $ given by Eq. (\ref{sigma}).

\section{Bell-type diagonal mixed states}

Let us analyze the three-photon mixed states of the quantum radiation field. With the help of the two-mode Fock basis $\{ \ket{3,0}, \ket{0,3}, \ket{2,1}, \ket{1,2}\}$, one can define the Bell-type basis as follows \cite{Ghiu-capitol}:
\beqa
\ket{\Psi^\pm}&=&\frac{1}{\sqrt 2}\; (\ket{3,0}\pm \ket{0,3}); \nonumber \\
\ket{\Phi^\pm}&=&\frac{1}{\sqrt 2}\; (\ket{2,1}\pm \ket{1,2}).\nonumber
\eeqa

The states $\ket{\Psi^\pm }$ are the well-known N00N states \cite{Dow, Bjork-2013}, while the other two states $\ket{\Phi^\pm}$ have recently been considered in Ref. \cite{Leonski}.

The Bell-type diagonal mixed states are defined by the following density operator
\beq
\rop =\alpha \proj{\Phi^+}{\Phi^+}+\beta \proj{\Phi^-}{\Phi^-}+\gamma \proj{\Psi^+}{\Psi^+}+\delta \proj{\Psi^-}{\Psi^-},
\label{roBD}
\eeq
where the parameters $\alpha $, $\beta $, $\gamma $, $\delta $ satisfy the normalization condition $\alpha +\beta + \gamma + \delta =1$.

The Chernoff quantum degree of polarization is obtained by using Eq. (\ref{mas-c}):
\beq
\pgen_{{\rm C}}(\rop )=1- \min_{s\in [0,1]}\, \left( \frac{1}{4}\right) ^{1-s}\left( \alpha^s+\beta^s+\gamma^s+\delta^s\right).
\label{w-c}
\eeq

Further, we want to evaluate the entanglement of a Bell-type diagonal mixed state in order to be able to compare the quantum degree of polarization with the entanglement. Concurrence is a useful tool for evaluating the entanglement of a mixed state \cite{Wootters-1998, Horodecki-2009}. The expression of the concurrence in the case of a Bell diagonal state reads \cite{Adesso}
\beq
C(\rop )=\max \{0, \lambda_1-\lambda_2-\lambda_3-\lambda_4\},
\label{conc-bd}
\eeq
where $\{ \lambda_1, \lambda_2, \lambda_3, \lambda_4\}$ are the eigenvalues $\alpha $, $\beta $, $\gamma $, $\delta $ in decreasing order, i.e. $\lambda_1\ge \lambda_2\ge \lambda_3\ge \lambda_4$.

\subsection{Werner-type state}

Here we focus our analysis on an important state of the class of Bell diagonal mixed states, namely
the Werner state \cite{Werner}:
\beq
\rop _W=a\, \proj{\Psi^-}{\Psi^-}+(1-a)\, \frac{1}{4}\, I,
\eeq
with $a$ a parameter that satisfies $a\in [0,1]$. The Werner state is defined by taking
\beqa
&&\alpha =\beta =\gamma =\frac{1-a}{4}, \nonumber \\
&&\delta =\frac{3\, a+1}{4}.
\eeqa
in the general expression of the Bell-type diagonal state (\ref{roBD}).

Firstly, we obtain the expression of the Chernoff quantum degrees of polarization by using Eq. (\ref{w-c}) \cite{Ghiu-2015, Ghiu-2016}:
\beq
\pgen_{{\rm C}}(\rop_W )=1-\frac{1}{4}\, \min_{s\in [0,1]} \, \left[ (1+3\, a)^s+3\, (1-a)^s\right] .
\eeq

Secondly, the concurrence of the Werner state is given by \cite{Alber}:
\beq
C(\rop _W)=\left\{ \begin{array}{lll}
0, & \mbox{if} & a\in \left[ 0,\frac{1}{3}\right] \\
\frac{1}{2}\, (3\, a-1), & \mbox{if} & a\in \bigg( \frac{1}{3},1\bigg] .
\end{array} \right.
\eeq

\begin{figure}
\centering
\includegraphics[width=8cm]{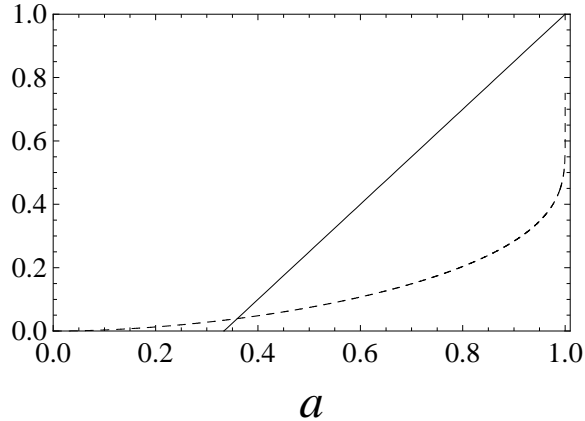}
\caption{The concurrence (solid) versus the Chernoff quantum degree of polarization (dashed) for the Werner state. The state with $\tilde a=0.3595871 $ fulfills the condition $C(\rop _W)=\pgen_{{\rm C}}(\rop_W )$. }
\label{werner-c}
\end{figure}

In Fig. \ref{werner-c} we plot the concurrence versus the Chernoff quantum degree of polarization. There is a special state with the property that the concurrence is equal to the Chernoff quantum degree of polarization, namely the state characterized by the parameter $\tilde a =0.3595871$.

\subsection{Bell-type diagonal states described by the parameter $x$}

Let us investigate an interesting class of Bell-type diagonal states, which is characterized by
the coefficients $\alpha $, $\beta $, $\gamma $, $\delta $ of the density operator (\ref{roBD}) that depend on a parameter $x\in [-1,1]$ as follows \cite{Ghiu-capitol}:
\beqa
\alpha (x)&=&\frac{1}{4}\, (1-x+x^2-x^3); \nonumber \\
\beta (x)&=&\frac{1}{4}\, (1-x-x^2+x^3); \nonumber \\
\gamma (x)&=&\frac{1}{4}\, (1+x-x^2-x^3); \label{st-2} \\
\delta (x)&=&\frac{1}{4}\, (1+x+x^2+x^3). \nonumber
\eeqa
We denote this density operator by $\rop (x)$.

The concurrence is obtained by using Eq. (\ref{conc-bd}):
\beq
C(\rop (x))
=\left\{ \begin{array}{lll}
\frac{1}{2}\, (-1-x+x^2-x^3), & \mbox{if} & x\in [-1,-0.544) \nonumber \\
0, & \mbox{if} & x\in [-0.544,0.544] \nonumber \\
\frac{1}{2}\, (-1+x+x^2+x^3), & \mbox{if} & x\in (0.544,1]
\end{array} \right.
\eeq

\begin{figure}
\centering
\includegraphics[width=8cm]{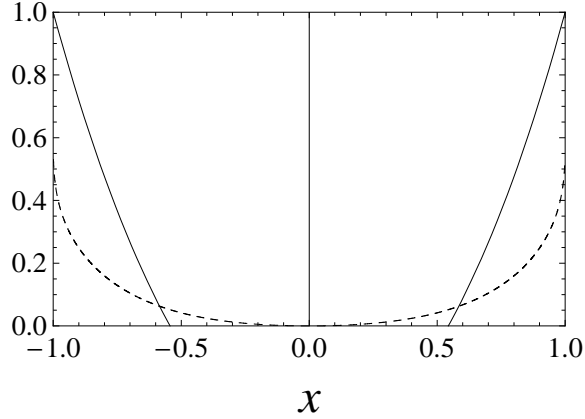}
\caption{A comparison between concurrence (solid) and the Chernoff degree of polarization (dashed) for the state defined by Eq. (\ref{st-2}). The two intersection points correspond to $\tilde x_-=- 0.584413$ and $\tilde x_+= 0.584413$.}
\label{st2-c}
\end{figure}

Figure \ref{st2-c} shows the behavior of the concurrence versus the quantum degree of polarization. The concurrence is equal to the Chernoff quantum degree of polarization in two cases, namely for the parameters $\tilde x = \pm 0.584413$.

\section{Conclusions}

The purpose of this paper was to make a comparison between entanglement and quantum degree of polarization for three-photon Bell-type diagonal states of the radiation field. We have considered the concurrence as a measure of entanglement and a distance-based quantity, namely the Chernoff bound for the quantum degree of polarization.

For a Werner-type state, one can notice from Fig. \ref{werner-c} that for $a<\tilde a$, the quantum degree of polarization is greater than the concurrence, while for $a>\tilde a$, the situation is the opposite: $\pgen (\rop_W )< C(\rop _W) $.

In the case of a Bell-type diagonal state depending an a parameter $x$, Fig. \ref{st2-c} presents the following behavior of the entanglement versus the quantum degree of polarization:
\[
{\rm C}(\rop )\; \;  {\rm versus} \; \; \pgen(\rop ): \left\{ \begin{array}{lll}
C(\rop (x))\ge \pgen (\rop(x) ), & \mbox{if} & x\in [ -1,\tilde x_-] \\
C(\rop (x))< \pgen (\rop(x) ), & \mbox{if} & x\in (\tilde x_-,\tilde x_+) \\
C(\rop (x))\ge \pgen (\rop(x) ), & \mbox{if} & x\in [\tilde x_+,1].
\end{array} \right.
\]
A similar comparison was found when using two other measures, Hilbert-Schmidt and Bures, for the quantum degree of polarization \cite{Ghiu-capitol}.

In conclusion, there is no dominance relation for the concurrence and the Chernoff quantum degree of polarization for all the domain of the parameters that define the states. Or, in other words, the entanglement and the quantum degree of polarization are incomparable measures, {\it i.e.} there exist states $\hat \rho_{1}$ and $\hat \rho_{2}$ such that $C(\rop _{1})<\pgen (\rop _{1})$, while $C(\rop _{2})>\pgen(\rop _{2})$.

\vspace{0.4cm}

\section*{Acknowledgements}
This work was supported by the funding agency CNCS-UEFISCDI of the Romanian Ministry of Research and Innovation through grant PN-III- P4-ID-PCE-2016-0794 within PNCDI III.


\begin{thebibliography}{99}

\bibitem{Nielsenbook} M. A. Nielsen and I. L. Chuang, {\it Quantum Computation and Quantum Information}, Cambridge University Press, Cambridge, U. K., 2000.

\bibitem{Sen} A. Sen (De), Current Science {\bf 112}, 1361 (2017).

\bibitem{Ghiu-2012} I. Ghiu,  J. Phys.: Conf. Ser. {\bf 338}, 012008 (2012).

\bibitem{Ghiu-2013} I. Ghiu, Phys. Scr. {\bf T153}, 014027 (2013).

\bibitem{Ghiu-math-2014} C. Ghiu and I. Ghiu, Cent. Eur. J. Math. {\bf 12}, 337 (2014).

\bibitem{Ghiu-phys-2014} I. Ghiu and C. Ghiu, Rep. Math. Phys. {\bf 73}, 49 (2014).

\bibitem{Manko-1-2014}  P. Adam, V. A. Andreev, I. Ghiu, A. Isar, M. A. Man'ko, and V. I. Man'ko, J. Russ. Laser Res.  {\bf 35}, 3 (2014).

\bibitem{Manko-2-2014}  P. Adam, V. A. Andreev, I. Ghiu, A. Isar, M. A. Man'ko, and V. I. Man'ko, J. Russ. Laser Res.  {\bf 35}, 427 (2014).

\bibitem{Olivares} S. Olivares, M. G. A. Paris, and R. Bonifacio, Phys. Rev. A {\bf 67}, 032314 (2003).

\bibitem{Cerf} N. J. Cerf, O. Kr\"{u}ger, P. Navez, R. F. Werner, and M. M. Wolf, Phys. Rev. Lett. {\bf 95}, 070501 (2005).

\bibitem{Marian-2013} I. Ghiu, P. Marian, and T. A. Marian, Phys. Scr. {\bf T153}, 014028 (2013).

\bibitem{Marian-pra-2013} P. Marian, I. Ghiu, and T. A. Marian, Phys. Rev. A {\bf 88}, 012316 (2013).

\bibitem{Modi} K. Modi, A. Brodutch, H. Cable, T. Paterek, and V. Vedral, Rev. Mod. Phys. {\bf 84}, 1655 (2012).

\bibitem{Scarani} C. Branciard, E. G. Cavalcanti, S. P. Walborn, V. Scarani, and H. M. Wiseman, Phys. Rev. A {\bf 85}, 010301 (2012).

\bibitem{Isar-2015} T. Mihaescu and A. Isar, Rom. J. Phys. {\bf 60}, 853 (2015).

\bibitem{Isar-eur-2017} A. Isar and T. Mihaescu, Eur. Phys. J. D {\bf 71}, 144 (2017).

\bibitem{Isar-proc-2017} T. Mihaescu and A. Isar, AIP Conference Proceedings {\bf 1796}, 020012 (2017).

\bibitem{Serban-2015} S. Suciu and A. Isar, Rom. J. Phys. {\bf 60}, 859 (2015).

\bibitem{Isar-2017} T. Mihaescu and A. Isar, Rom. J. Phys. {\bf 62}, 107 (2017).

\bibitem{Bjork-2005} A. B. Klimov, L. L. S\'anchez-Soto, E. C. Yustas, J. S\"{o}derholm, and G. Bj\"{o}rk, Phys. Rev. A {\bf 72}, 033813 (2005).

\bibitem{Ghiu-2010} I. Ghiu, G. Bj\"{o}rk, P. Marian, and T. A. Marian, Phys. Rev. A {\bf 82}, 023803 (2010).

\bibitem{Bjork-2010} G. Bj\"{o}rk, J. S\"{o}derholm, L. L. S\'anchez-Soto, A. B. Klimov, I. Ghiu,
P. Marian, and T. A. Marian, Opt. Comm. {\bf 283}, 4440 (2010).

\bibitem{Silva} K. Nogueira, J. B. R. Silva, J. R. Goncalves, and H. M. Vasconcelos, Phys. Rev. A  {\bf 87}, 043821 (2013).

\bibitem{Ghiu-capitol} I. Ghiu, {\it Polarization and Entanglement of Three-Photon Bell-Diagonal Mixed States},
pp. 99, in {\it New Developments in Quantum Optics Research}, N. Stewart (Ed.), Nova
Science Publishers, 2015.

\bibitem{Chernoff} H. Chernoff, Ann. Math. Stat. {\bf 23}, 493 (1952).

\bibitem{Nussbaum} M. Nussbaum and A. Szko\l a, Ann. Stat. {\bf 37}, 1040 (2009).

\bibitem{Acin} K. M. R. Audenaert, J. Calsamiglia, R. Mu\~noz-Tapia, E. Bagan,
Ll. Masanes, A. Acin, and F. Verstraete, Phys. Rev. Lett. {\bf 98}, 160501 (2007).

\bibitem{Kargin} V. Kargin, Ann. Stat. {\bf 33}, 959 (2005).

\bibitem{Auden} K. M. R. Audenaert, M. Nussbaum, A. Szko\l a, and F. Verstraete,
Comm. Math. Phys. {\bf 279}, 251 (2008).

\bibitem{Lehner} J. Lehner, U. Leonhard, and H. Paul, Phys. Rev. A {\bf 53}, 2727 (1996).

\bibitem{Dow} J. P. Dowling, Cont. Phys. {\bf 49}, 125 (2008).

\bibitem{Bjork-2013} S. Shabbir, M. Swillo, and G. Bj\"{o}rk, Phys. Rev. A {\bf 87}, 053821 (2013).

\bibitem{Leonski} A. Kowalewska-Kudlaszyk and W. Leonski, J. Opt. Soc. Am. B {\bf 31}, 1290 (2014).

\bibitem{Wootters-1998} W. K. Wootters, Phys. Rev. Lett. {\bf 80}, 2245 (1998).

\bibitem{Horodecki-2009} R. Horodecki, P. Horodecki, M. Horodecki, and K. Horodecki, Rev. Mod. Phys. {\bf 81}, 865 (2009).

\bibitem{Adesso} T. R. Bromley, M. Cianciaruso, R. Lo Franco, and G. Adesso, J. Phys. A: Math. Theor. {\bf 47}, 405302 (2014).

\bibitem{Werner}R. F. Werner, Phys. Rev. A {\bf 40}, 4277 (1989).

\bibitem{Ghiu-2015} I. Ghiu, C. Ghiu, and A. Isar, Proc. Romanian Acad. A {\bf 16}, 499 (2015).

\bibitem{Ghiu-2016} I. Ghiu and A. Isar, Rom. J. Phys. {\bf 61}, 768 (2016).

\bibitem{Alber} M. Ali, M. R. P. Rau, and G. Alber, Phys. Rev. A {\bf 81}, 042105 (2010).



\end{thebibliography}
\end{document}